\documentclass[twocolumn,preprintnumbers,showpacs,prl,amsmath,amssymb]{revtex4}
\usepackage{graphicx}% Include figure files
\usepackage{dcolumn}% Align table columns on decimal point
\usepackage{bm}% bold math

\begin{document}

%\preprint{APS/123-QED}

\title{Low temperature studies of the excited-state structure of negatively-charged Nitrogen-Vacancy color centers in diamond}% Force line breaks with \\

\author{A.~Batalov$^{1}$, V.~Jacques$^{1,\ast}$, F. Kaiser$^{1}$, P. Siyushev$^{1}$, P.~Neumann$^{1}$, L.~J.~Rogers$^{2}$, R.~L.~McMurtrie$^{2}$, N.~B.~Manson$^{2}$, F.~Jelezko$^{1}$ and J.~Wrachtrup$^{1}$}

%\email{v.jacques@physik.uni-stuttgart.de}

\affiliation{$^{1}$3. Physikalisches Institut, Universit$\ddot{\rm a}$t Stuttgart, 70550 Stuttgart, Germany\\
$^{2}$Laser Physics Centre, Australian National University, Canberra,
ACT 0200, Australia}

\altaffiliation{v.jacques@physik.uni-stuttgart.de}

\date{\today}

\begin{abstract}
We report a study of the $^{3}$E excited-state structure of single negatively-charged nitrogen-vacancy (NV) defects in diamond, combining resonant excitation at cryogenic temperatures and optically detected magnetic resonance. A theoretical model is developed and shows excellent agreement with experimental observations. Besides, we show that the two orbital branches associated with the $^{3}$E excited-state are averaged when operating at room temperature. This study leads to an improved physical understanding of the NV defect electronic structure, which is invaluable for the development of diamond-based quantum information processing.
\end{abstract}

\maketitle
\begin{figure}[t]
 \centerline{\includegraphics[width=8.5cm]{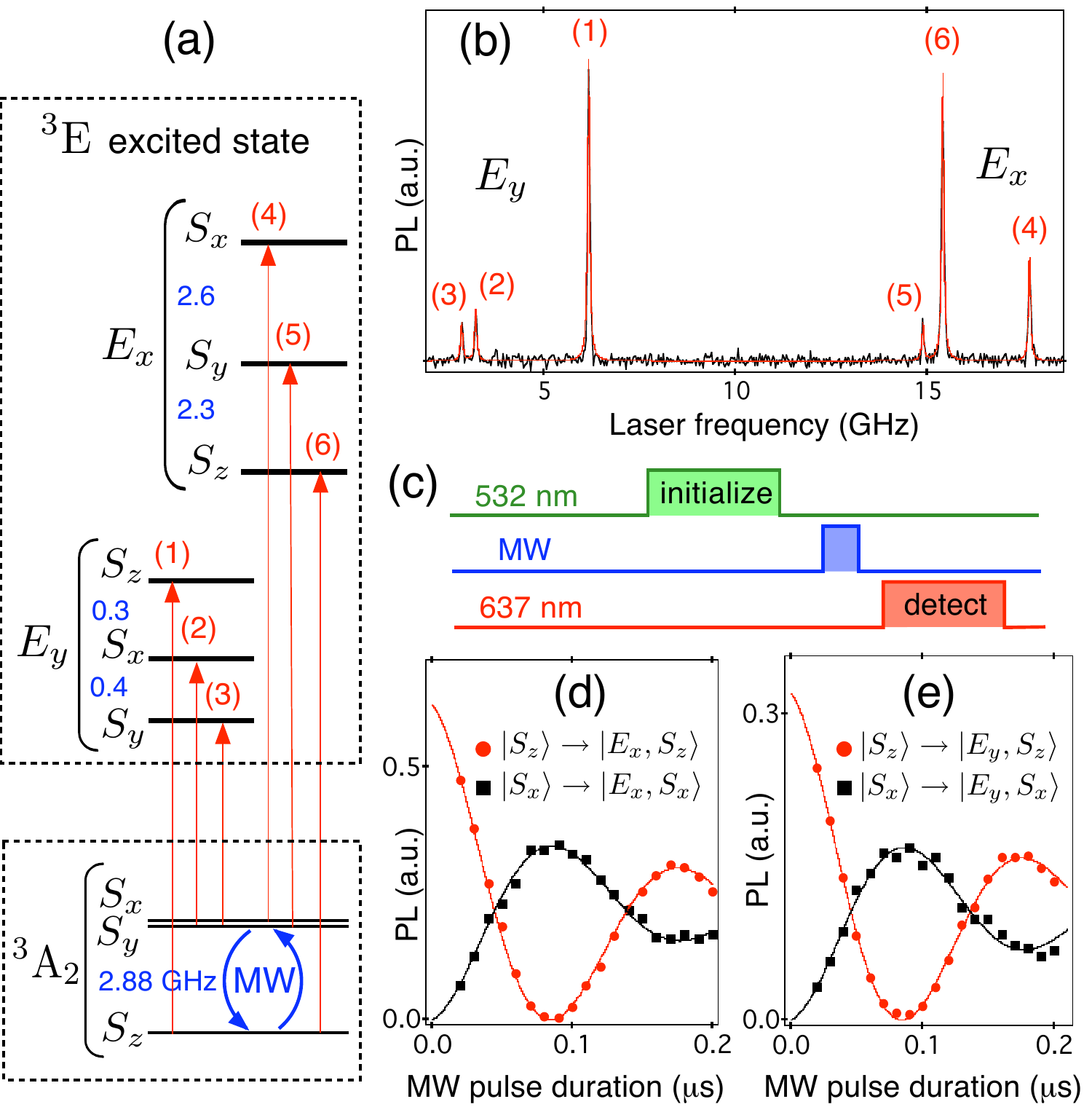}}
    \caption{(color online). (a)-Energy-level diagram of a single NV defect. (b)-Excitation spectrum of a single NV defect presenting six resonances, which correspond to optical transitions linking identical spin sublevels. Solid line is data fitting using Lorentzian functions. Energy splittings between each excited-state spin sublevels for this NV center are written in blue in (a) (not in scale). (c)-Pulse sequence used to measure single-electron spin Rabi oscillations (see main text). (d) and (e)-Rabi nutations measured using resonant optical transitions for the spin read-out in the upper branch $E_{x}$ (d) and in the lower branch $E_{y}$ (e). A $\pi$ phase shift of the Rabi nutation allows to discriminate for each orbital branch between optical transitions linking $S_{z}$ or $S_{x,y}$ spin sublevels.}
    \label{Fig1}
    \end{figure}

\indent Coupling between flying and stationary qubits is one of the crucial requirements for scalable quantum information processing~\cite{Duan_Nature2001,Barrett_PRA2005}. Among many quantum systems including single atoms~\cite{RevueIons_Nature2008} and semiconductor quantum dots~\cite{Imamoglu_PRL1999}, the negatively-charged nitrogen-vacancy (NV) color center in diamond is a promising solid-state candidate for realizing such interface, owing to long spin coherence time of their spin states~\cite{Gupi_NatMat2009} and availability of a strong optical transition~\cite{Gruber_Science1997}. Nevertheless, even though NV defects have been intensively studied during the last decades, the excited-state structure as well as the dynamics of excitation-emission cycles are surprisingly not yet fully understood. This knowledge is however of crucial importance for the realization of long-distance entanglement protocols based on coupling of spin-state to optical transitions~\cite{Moehring_Nature2007,Olmschenk_Science2009,Lukin_PRL2006}.\\
\indent In this Letter, we report a study of the excited-state structure of single NV defects as a function of local strain, combining resonant excitation at cryogenic temperatures and optically detected magnetic resonance (ODMR). Besides, we show that the two orbital branches associated with the $^{3}$E excited-state are averaged at room temperature. A theoretical model is developed and shows excellent correspondence with experimental observations. \\
%This work provides significant insight into the structure of the emitting excited state, which is invaluable for future realization of long-distance entanglement protocols.\\
\indent The NV color center in diamond consists of a substitutional nitrogen atom (N) associated with a vacancy (V) in an adjacent lattice site, giving a defect with $\rm C_{3v}$ symmetry. For the negatively-charged NV color center addressed in this study, the ground state is a spin triplet $^{3} \rm A_{2}$~\cite{Manson_PRB2006,Gali_PRB2008,Newton_PRB2009}. Spin-spin interaction splits ground state spin sublevels by $2.88$ GHz into a spin singlet $S_{z}$, where $z$ corresponds to the NV symmetry axis, and a spin doublet $S_{x},S_{y}$ (see Fig.~\ref{Fig1}(a)). The excited state $^{3} \rm E$ is also a spin triplet, associated with a broadband photoluminescence emission with zero phonon line (ZPL) around $637$ nm ($1.945$ eV). Besides, the $^{3} \rm E$ excited state is an orbital doublet, which degeneracy is lifted by non-axial strain into two orbital branches, $E_{x}$ and $E_{y}$, each orbital branch being formed by three spin states $S_{x}$, $S_{y}$ and $S_{z}$ (see Fig.~\ref{Fig1}(a))~\cite{Tamarat_NJP2008}. As optical transitions $^{3} \rm A_{2}\rightarrow ^{3}$E are spin-conserving, excitation spectra of single NV color centers might show six resonant lines, corresponding to transitions between identical spin sublevels. \\
\indent The order of other energy levels is still under debate but it is now well established that at least one metastable state $^{1}\rm A_{1}$ is lying between the ground and excited triplet states~\cite{Rogers_CondMat2008}. Non-radiative inter-system crossing to the $^{1}\rm A_{1}$ state is strongly spin selective as the shelving rate from the $S_{z}$ sublevel is much smaller than those from $S_{x}$ and $S_{y}$. Furthermore, the metastable state decays preferentially towards the ground state spin level $S_{z}$, leading to a strong spin polarization into that state after a few optical excitation-emission cycles. Thereby, optical transitions linking $S_{x}$ or $S_{y}$ sublevels are non cycling transitions. Experimental investigation of the excited-state structure then requires the use of a microwave excitation resonant with the ground state transition at $2.88$ GHz, in order to maintain a non-zero time-averaged population within each of the ground state spin sublevels~\cite{Tamarat_NJP2008}. \\
\indent Spectroscopic investigations require good spectral stability of NV defects. Using low nitrogen concentration type IIa natural diamond, Fourier-transform limited emission have been recently reported at cryogenic temperature~\cite{Tamarat_PRL2006,Batalov_PRL2008}. In the present study, we investigate native single NV defects in an ultra-pure synthetic type IIa diamond crystal prepared using a microwave assisted chemical vapor deposition process. Within such sample, where the nitrogen concentration is below 1ppb ($<1.7 \times 10^{14} \rm cm^{-3}$) and the NV defect concentration smaller than $10^{10}\rm cm^{-3}$, all single NV defects have shown perfect spectral stability.\\
\indent NV defects are addressed using confocal microscopy at cryogenic temperatures ($T\approx 4$ K). A tunable laser diode is used to excite resonantly the NV centers on their ZPL. The red-shifted photoluminescence (PL) between 650 nm and 750 nm is detected in a confocal arrangement and used to monitor excitation spectra by sweeping the laser diode frequency. In addition, microwaves (MW) resonant with the ground state spin transition are applied via a copper microwire located close to the NV defects.\\
A typical excitation spectrum of a single NV center is depicted in Fig.~\ref{Fig1}(b). As expected, three transitions are observed for each excited-state orbital $E_{x}$ and $E_{y}$, corresponding to transitions between identical spin sublevels. Unambiguous assignment of optical transitions linking $S_{z}$ or $S_{x},S_{y}$ spin sublevels is made by measuring single electron spin Rabi oscillations using the pulse sequence shown in Fig.~\ref{Fig1}(c). The NV center is first initialized into the ground state $S_{z}$ sublevel using an optical pulse of duration $3 \ \mu$s at the wavelength $\lambda=532$ nm. A microwave pulse at $2.88$ GHz is then applied and the spin-state is finally read-out by measuring the PL intensity using a $1 \ \mu$s red laser pulse resonant with a given transition. Depending on the optical transion used for the spin read-out ($S_{z}$ or $S_{x},S_{y}$), the Rabi nutation shows a $\pi$ phase shift, allowing to discriminate between optical transitions linking different spin sublevels (see Fig.~\ref{Fig1}(d)-(e)).\\
    \begin{figure}[b]
 \centerline{\includegraphics[width=8cm]{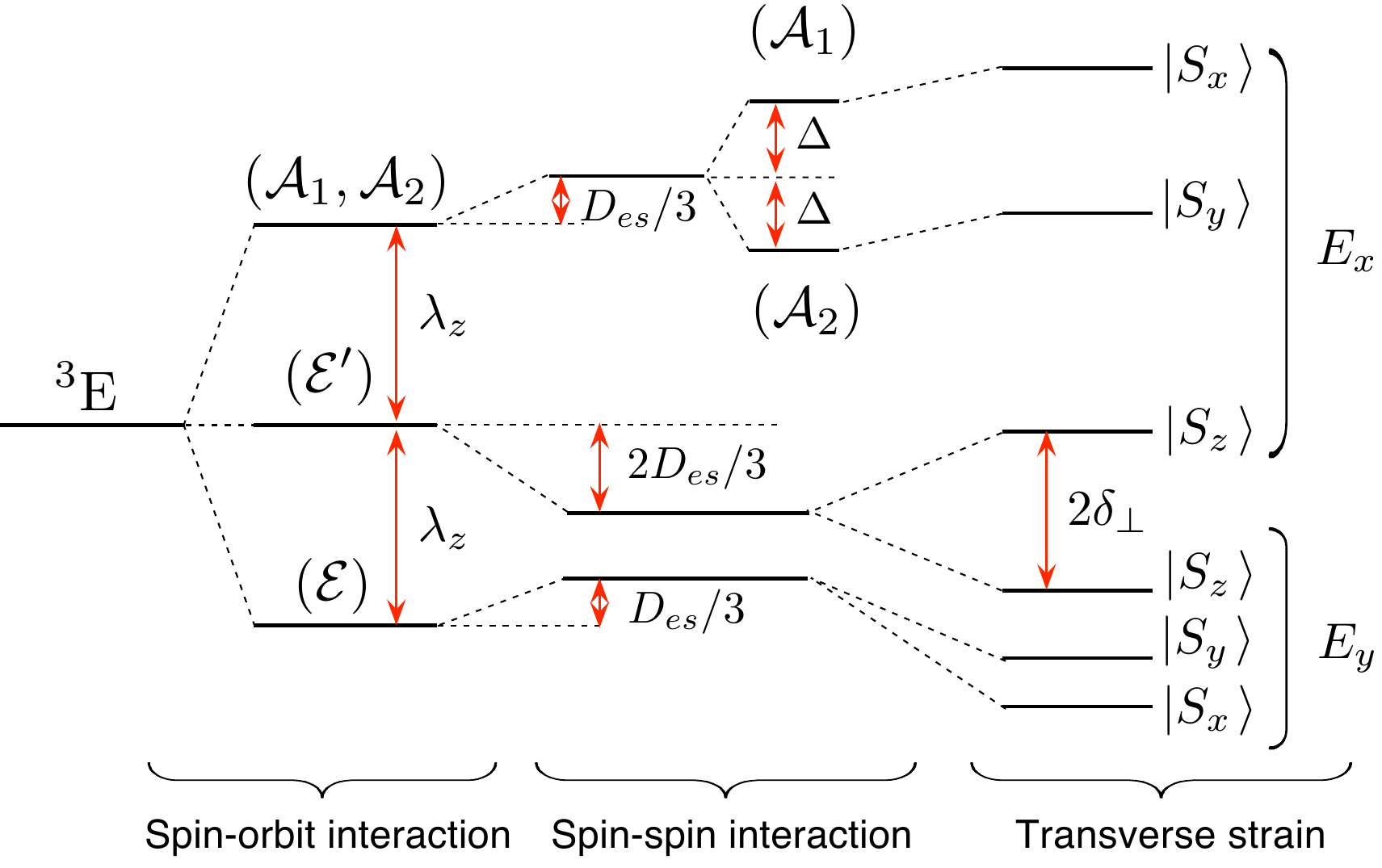}}
    \caption{(color online). Schematic diagram of the excited-state energy levels including spin-orbit interaction, spin-spin interaction and transverse local strain.  Note that $1$ GPa external stress gives approximately $10^{3}$ GHz splitting~\cite{Davies}.}
    \label{Fig2}
    \end{figure}
The energy level scheme presented in Fig.~\ref{Fig1}(a) is only valid for a particular local strain of the diamond lattice in the vicinity of the NV defect. Indeed, non-axial local strain strongly modifies the energy difference between the two orbitals $E_{x}$ and $E_{y}$, as well as the relative position of the spin sublevels. Following a model previously developed in Ref.~\cite{Tamarat_NJP2008}, we investigate the position of the energy levels as a function of the local strain. The Hamiltonian of the $^{3}$E excited-state is given by
\begin{equation}
H=H_{0}+H_{so}+H_{ss}+H_{str} \ ,
\label{Hamilto}
\end{equation}
where $H_{0}$ is the dominant term giving the energy of $1.945$ eV with respect to the $^{3}\rm A_{2}$ ground state, $H_{so}$ is the spin-orbit coupling, $H_{ss}$ the spin-spin interaction, and $H_{str}$ the perturbation resulting from local strain. The axial spin-orbit interaction $\lambda_{z}\hat{L}_{z}\hat{S}_{z}$ splits the $^{3}$E excited-state into three twofold degenerate levels, noted $(\mathcal{E})$, $(\mathcal{E}^{\prime})$ and $(\mathcal{A}_{1},\mathcal{A}_{2})$ according to the irreducible representations of $C_{3v}$ symmetry (see Fig.~\ref{Fig2}). The transverse spin-orbit interaction $\lambda_{x,y}$ is weak ($\lambda_{x,y}= 0.2$ GHz~\cite{Tamarat_NJP2008}) and can be neglected at present. Spin-spin interactions displace all the latter levels in a similar fashion as in the ground state by the energy $D_{es}(\hat{S}_{z}^{2}-2/3)$. Besides $H_{ss}$ lifts the degeneracy of the $\mathcal{A}_{1}$ and $\mathcal{A}_{2}$ states by $\pm \Delta$~\cite{Lenef_PRB1996}. The fine structure at zero strain is therefore given by the three parameters $\lambda_{z}$, $D_{es}$ and $\Delta$ (see Fig.~\ref{Fig2}).\\
\indent The effect of local strain can be described using the Hamiltonian $H_{str}=\delta_{x}\hat{V}_{x}+\delta_{y}\hat{V}_{y}+\delta_{z}\hat{V}_{z}$, where $\delta_{i}$ and $\hat{V_{i}}$ represent respectively the strain parameter and an orbital operator in the $i$ direction~\cite{Lenef_PRB1996}. Axial strain $\delta_{z}$ gives rise to a linear shift of all energy levels and is not considered further. Transverse strains, $\delta_{x}$ and $\delta_{y}$, split the energy levels into two spin triplets, associated with the two orbital branches $E_{x}$ and $E_{y}$. The energy splitting is given by $\pm \delta_{\bot}$, where $\delta_{\bot}\propto (\delta_{x}^{2}+\delta_{y}^{2})^{\frac{1}{2}}$ (see Fig.~\ref{Fig2})~\cite{Tamarat_NJP2008}.\\
\indent Excitation spectra of several single NV defects have been measured, corresponding to different values of the transverse strain $\delta_{\bot}$ (see Fig.~\ref{Fig3}(a)-(b)). From low strain experimental data, we estimate the zero-strain parameters to be $\lambda_{z}=5.3$ GHz, $D_{es}=1.42$ GHz and $\Delta=1.55$ GHz. Using such measurements, a calculation of the eigenenergies of the Hamiltonian described by equation~(\ref{Hamilto}) is performed for different strengths of the transverse strain $\delta_{\bot}$, including the effect of transverse spin-orbit interaction $\lambda_{x,y}= 0.2$ GHz~\cite{Tamarat_NJP2008}. All measurements are in excellent agreement with this calculation, as depicted in figure~\ref{Fig3}(b). \\
\indent With such results, we have a complete physical understanding of the NV defect dynamics in terms of cycling spin conserving transitions and spin-flip transitions.\\
\begin{figure}[t]
 \centerline{\includegraphics[width=8.5cm]{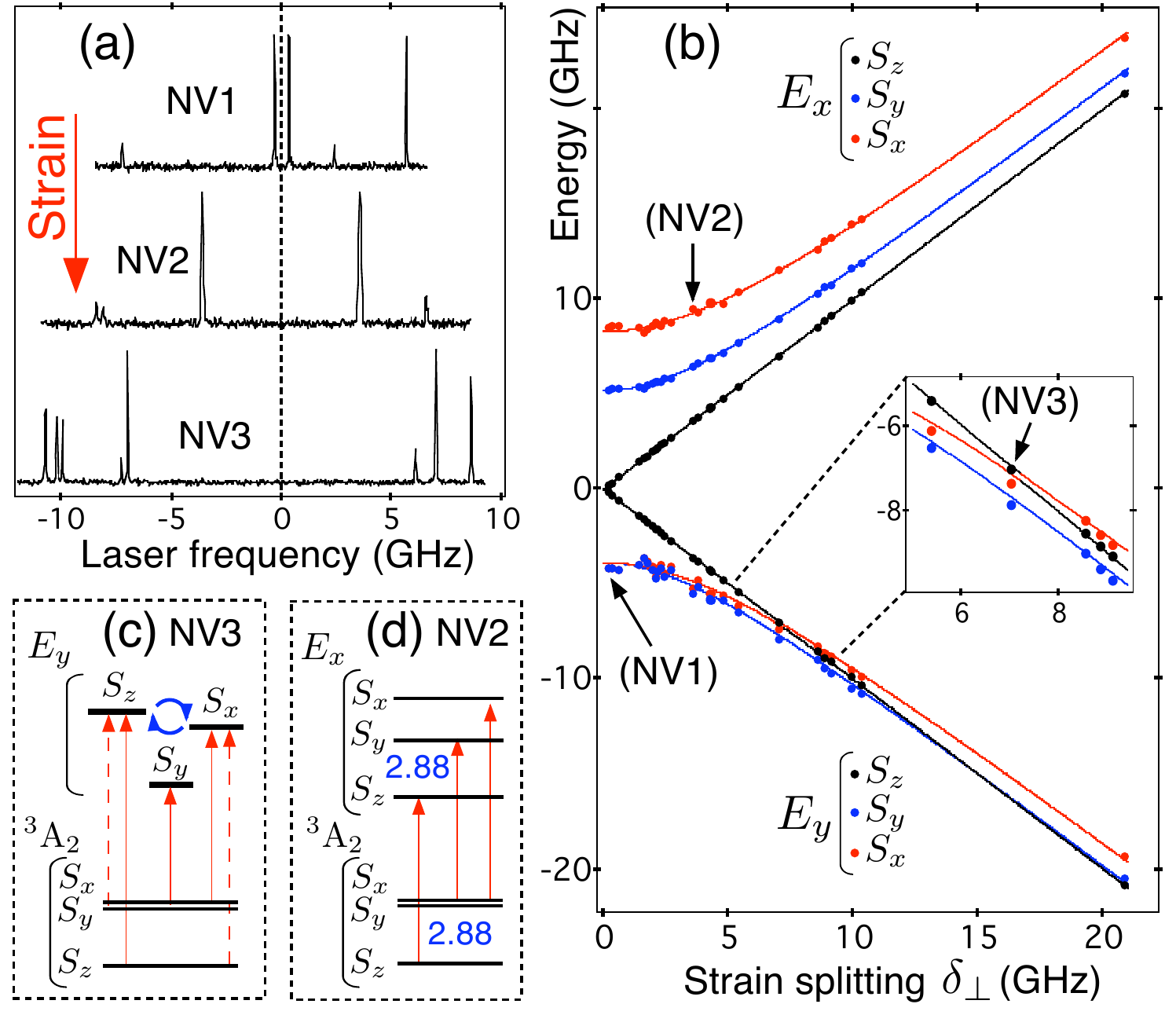}}
    \caption{(color online). (a)-Excitation spectra recorded for different single NV defects denoted NV1, NV2 and NV3. (b)-Excited-state energy structure as a function of the transverse strain $\delta_{\bot}$ for $27$ single NV defects. Solid lines correspond to the calculation without any free parameter and the inset is a zoom on the lower branch for a strain corresponding to an avoided crossing between $ \left|E_{y},S_{z}\right\rangle$ and $ \left|E_{y},S_{x}\right\rangle$ spin sublevels. At this avoided crossing, the solid lines exchange colors for clarity purpose. (c)-Energy level scheme of the lower branch for the defect NV3, showing three spin conserving transitions (solid arrows) and two spin-flip transitions (dashed arrows), appearing when the states $\left|E_{y},S_{z}\right.\rangle$ and $\left|E_{y},S_{x}\right.\rangle$ are crossing. (d)-Energy level scheme of the upper branch for the defect NV2. The energy splitting between the states $\left|E_{x},S_{z}\right.\rangle$ and $\left|E_{x},S_{y}\right.\rangle$ is equal to $2.88$ GHz, leading to a repumping effect of the  $\left|S_{z} \right\rangle \rightarrow \left|E_{x},S_{z}\right\rangle$ optical transition.}
\label{Fig3}
\end{figure} 
In the lower branch $E_{y}$, spin sublevels are strongly mixed by local strain, resulting in non-cycling transitions, as spin-flip can occur either by radiative decay or by non-radiative decay through the metastable state. Thus, the lower branch cannot be observed in excitation spectra without applying MW. Moreover, for a particular strain corresponding to an avoided crossing between $ \left|E_{y},S_{z}\right\rangle$ and $ \left|E_{y},S_{x}\right\rangle$ spin sublevels (see inset of Fig.~\ref{Fig3}(b)), the mixing becomes so important that fully allowed spin-flip optical transitions can be observed. This situation corresponds to the excitation spectrum measured for the defect NV3 (see Fig.~\ref{Fig3}(a) and (c)) where two additional spin-flip transitions are observed in the lower branch. We note that the same effect occurs at higher strain when $\left|E_{y},S_{z}\right\rangle$ and $ \left|E_{y},S_{y}\right\rangle$ spin sublevels are crossing. Such spin-flip transitions can be used for single-spin high-speed coherent optical manipulation through $\Lambda$-based scheme~\cite{Santori_OptExp2006,Santori_PRL2006}.\\
\indent In the upper branch $E_{x}$, the situation is completely different as no crossing occurs between different spin-sublevels. When the MW is not applied, a single resonant line remains visible in the excitation spectrum, corresponding to the cycling transition $\left|S_{z} \right\rangle \rightarrow \left|E_{x},S_{z}\right\rangle$. Nonetheless, for most of the studied NV defects this resonance is observed to be several times weaker without applying MW. This is explained by a remaining small mixing of the excited-state spin sublevels through non-axial spin orbit interaction $\lambda_{x,y}$~\cite{Manson_PRB2006}. Remarkably, for a specific local strain, the optical transition $\left|S_{z} \right\rangle \rightarrow \left|E_{x},S_{z}\right\rangle$ is however found to be much stronger without applying MW. This situation is described in Fig.~\ref{Fig3}(d), corresponding to the excitation spectrum of defect NV2 shown in Fig.~\ref{Fig3}(a). At this particular strain the energy splitting between $\left|E_{x},S_{z}\right.\rangle$ and $\left|E_{x},S_{y}\right.\rangle$ is equal to $2.88$ GHz, like in the ground state. In such a configuration, if a spin-flip occurs during $\left|S_{z} \right\rangle \rightarrow \left|E_{x},S_{z}\right\rangle$ transitions, the NV defect is efficiently repumped in the $S_{z}$ ground state through a resonant $\left|S_{y} \right\rangle \rightarrow \left|E_{x},S_{y}\right\rangle$ optical transition followed by non-radiative decay to the metastable state responsible for spin-polarization of the NV defect. Note that local strain can be tuned by applying electric field through electrodes deposited on the diamond surface, allowing to externally control conditions of spin-conserving cycling transitions or spin-flip transitions for $\Lambda$ scheme~\cite{Tamarat_NJP2008}.\\
\indent Excited-state spectroscopy of single NV defect using ODMR techniques at room temperature has recently underlined an excited-state electron spin resonance (ESR) around $1.4$ GHz~\cite{Afshalom_PRL2008,Neumann_QuantPh2008}. Surprisingly, low temperature excitation spectra described above do not show any signature of such an energy splitting between excited-state spin sublevels, neither in the upper branch nor in the lower branch. In order to understand this observation, ODMR spectra are measured for the same single NV defect at different temperatures by sweeping the MW frequency. As depicted in Fig.~\ref{Fig4}(a), when the experiment is performed close to room temperature, the well-known ground state ESR is detected at $2.88$ GHz, and the excited-state ESR around $1.4$ GHz is observed, as previously reported. Performing the same experiment at low temperature (T$=6$ K) shows that the excited-state ESR fully disappears. The evolution of the excited-state ESR contrast as a function of temperature is depicted in Fig.~\ref{Fig4}(b), showing a strong decrease around $T=150$ K. 
\begin{figure}[t]
\centerline{\includegraphics[width=8.5cm]{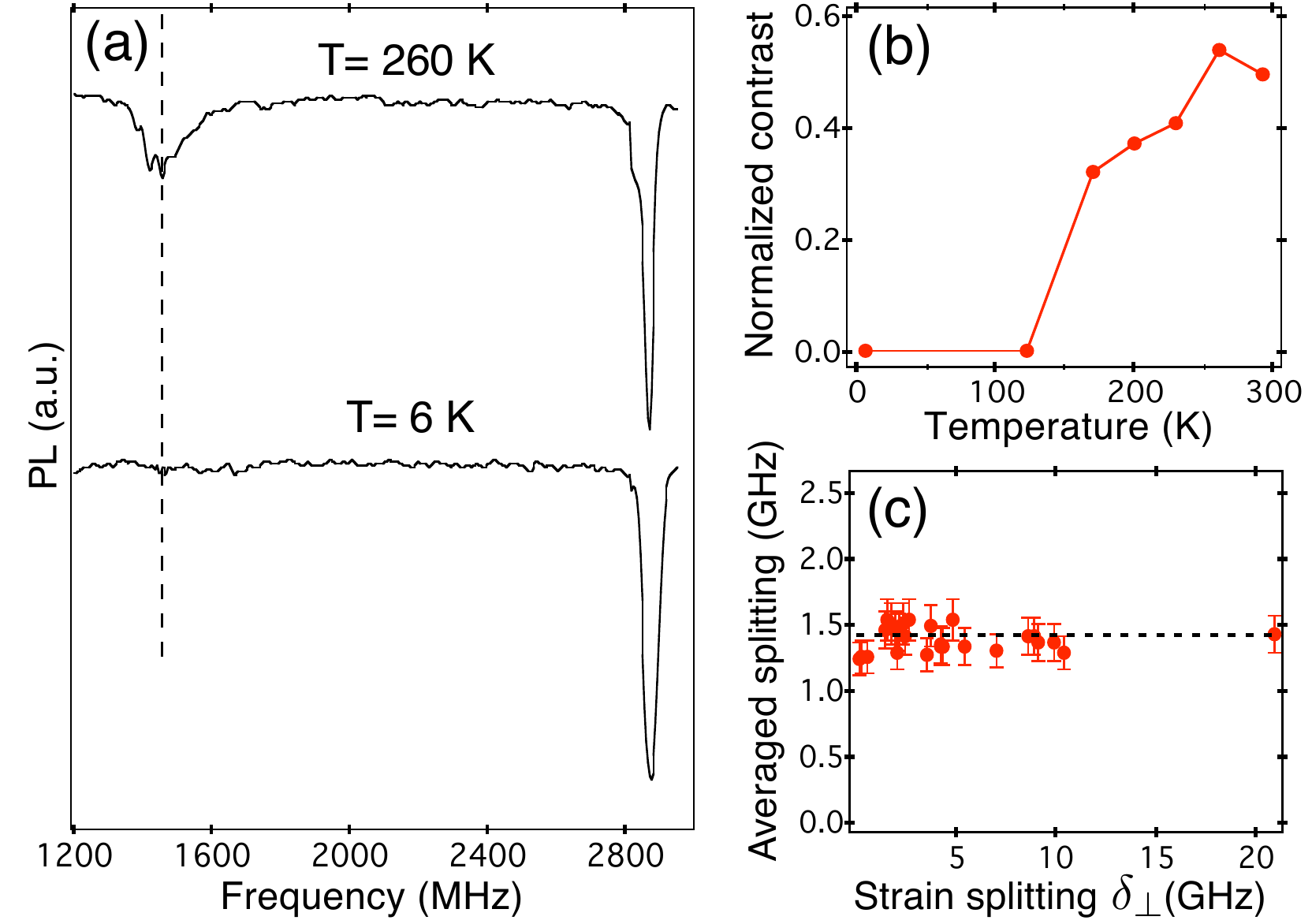}}
\caption{(color online). (a)-ODMR spectra of a single NV color center recorded at $T=260$ K and at $T=6$ K. (b)-Contrast of the excited-state ESR as a function of temperature. The contrast is normalized to the integral of the ground state ESR at $2.88$ GHz. (c)-Energy splitting between the averaged energy of $S_{z}$ and $S_{x,y}$ sublevels of the two orbital branches as a function of strain. Data points are inferred using the measurements depicted in Fig.~\ref{Fig3}(b). The black dashed line corresponds to $D_{es}=1.42$ GHz.}
\label{Fig4}
\end{figure}

This observations are explained by considering that the two excited-state branches are averaged when operating at room temperature~\cite{Rogers_ESRAver}. Room temperature studies have indeed shown that the isotropic g-factor associated with the excited-state ESR at $1.4$ GHz is similar to the ground state g-factor ($g \approx 2$)~\cite{Afshalom_PRL2008,Neumann_QuantPh2008}, indicating that the orbital angular momentum does not play a significant role in the excited state at room temperature. We now assess the expected value of the energy splitting between $S_{z}$ and $S_{x},S_{y}$ excited-state spin sublevels by averaging over the orbital branches. Such an orbital averaging quenches both the effect of spin-orbit splitting $\lambda_{z}$ and the effect of spin-spin interaction $\Delta$, which splits $\mathcal{A}_{1}$ and $\mathcal{A}_{2}$ sublevels~\cite{Lenef_PRB1996} (see Fig.~\ref{Fig2}). The averaged energy splitting between $S_{z}$ and $S_{x},S_{y}$ excited-state spin sublevels is then only determined by the parameter $D_{es}$, which has a value of $1.42$ GHz. This averaged value can be checked using the measurements shown in Fig.~\ref{Fig3}(b), by estimating the energy splitting between averaged $S_{z}$ and $S_{x},S_{y}$ spin sublevels. Such splitting is found around $1.4$ GHz for the whole range of considered local strains, as depicted in Fig.~\ref{Fig4}(c). This is consistent with the value of $D_{es}$ and, hence, the averaging process explains why an excited-state ESR is detected around $1.4$ GHz at room temperature. We note that for very large strain, which leads to a lowering of $C_{3v}$ symmetry, the spin-spin interaction becomes $H_{ss}=D_{es}(\hat{S}_{z}^{2}-2/3)+E_{es}(\hat{S}_{x}^{2}-\hat{S}_{y}^{2})$ where $E_{es}$ is proportional to the transverse strain. In such conditions, the averaged energy splitting is then given by $D_{es}\pm E_{s}$. In case of large strain, the orbital averaging then accounts for previously reported observations of a splitted excited-state ESR at room temperature~\cite{Afshalom_PRL2008,Neumann_QuantPh2008}.\\
\indent The physical process responsible for orbital averaging, which remains under question, will be addressed in future work. This process, which could induce flips between the two orbitals at a frequency much higher than the radiative lifetime without altering the spin projection, could be for example a phonon-mediated process~\cite{Rogers_ESRAver}. Similar process would also account for why ESR is not observed in the $^{2}$E gound state of the neutral NV defect~\cite{Newton_PRB2008}. \\
\indent Summarizing, we have reported a study of the negatively-charged NV defect excited-state structure as a function of local strain. The present work gives significant insights into the NV defect electronic structure, which is invaluable for the development of diamond-based quantum information processing.\\
\indent The authors are grateful to R.~Kolesov and G.~Balasubramanian for fruitful discussions. We acknowledge D.~Twitchen and M.~Markham from Element6 (UK) for providing ultra-pure diamond samples. This work is supported by the European Union (QAP, EQUIND, NEDQIT), Deutsche Forschungsgemeinschaft (SFB/TR21) and Australian Research Council. V. J. acknowledges support by the Humboldt Foundation.

 \end{document}